*Designing everyday automation with well-being in mind*

Holger Klapperich, Alarith Uhde & Marc Hassenzahl

ONLINE FIRST

Springer

ORIGINAL ARTICLE

# Designing everyday automation with well-being in mind

Holger Klapperich [1] · Alarith Uhde [1] · Marc Hassenzahl [1]

**Abstract**
Nowadays, automation not only permeates industry but also becomes a substantial part of our private, everyday lives. Driven by the idea of increased convenience and more time for the "important things in life," automation relieves us from many daily chores—robots vacuum floors and automated coffee makers produce supposedly barista-quality coffee on the press of a button. In many cases, these offers are embraced by people without further questioning. However, while we save time by delegating more and more everyday activities to automation, we also may lose chances for enjoyable and meaningful experiences. In two field studies, we demonstrate that a manual process has experiential benefits over more automated processes by using the example of coffee-making. We present a way to account for potential experiential costs of everyday automation and strategies of how to design interaction with automation to reconcile experience with the advantages of a more and more powerful automation.

## 1 Introduction

Since the Stone Age, humans make tools. Over the centuries, these tools became more and more powerful and efficient, allowing for new capabilities as well as easing the strains of physical labor. In the second half of the eighteenth century, toolmaking turned gradually into automation, with spinning mules and looms powered by horses, water, and steam. This development was a transfer of labor from humans to machines. However, this was not done primarily to humanize work, but to produce goods more efficiently and profitably. What seemed to disappear from view was that humans make use of the material world not only to survive, but also to carry out projects of self-realization [1]. Work is not only about efficiency, it is also about personal growth and development. Despite this insight, automation made its way from factories into homes and our daily lives. For example, the fully automated toaster supports the busy "housewife" since 1926. Unsurprisingly, not only automation itself, but also related values and beliefs have entered the home: Efficiency became as important in everyday private life as it is in business.

However, efficiency may not be especially crucial to domestic comfort. Pillan and Colombo [2] suggested to understand the domestic as "complex context where people experience and express some of the most important and basic dimensions of human existence" ([2], p. 2584). For them, a mixture of "identity," "emotions," and "social relations" represents the factors crucial for well-being at home. Shove et al. [3] explain the success of everyday automation and efficiency by distinguishing two types of technologies: "While some innovations add to the pressures of daily life […], others, like so-called "labor-saving technologies," promise to release time for competing pursuits" ([3], p. 23). Evans [4] goes a step further by distinguishing "donative designs," that is, products we like to spend time on, from "compressive designs," that is, products which save time. Both categories are related: "compressive and donative […] complement each other, neither would make a great deal of sense without the other" ([4], p. 18). A good example for this is Grosse-Hering et al.'s [5]

✉ Holger Klapperich
holger.klapperich@uni-siegen.de

Alarith Uhde
alarith.uhde@uni-siegen.de

Marc Hassenzahl
marc.hassenzahl@uni-siegen.de

1 'Ubiquitous Design', Experience & Interaction, University Siegen, Kohlbettstraße 15, 57072 Siegen, Germany

strategy to deliberately prolong and emphasize meaningful moments of an activity and to downplay less important aspects. *Juicy-Mo* is a device to juice fruits. Its design deliberately prolongs the joyful elements (i.e., juicing, preparing) and shortens the not so joyful ones (i.e., cleaning).

While automation has many advantages (e.g., time saved, better and more predictable results), it also has well-known negative side effects, including alienation, deskilling, and overreliance. Aporta and Higgs [6], for example, studied how the Inuit lost their skills in wayfinding after GPS-technology became widely used. Originally, wayfinding skills were passed from generation to generation—they had been integral to survival. This ended with the rise of electronic navigation systems. As a consequence, the Inuit and their survival under harsh conditions became highly dependent on GPS, since they could not fall back to practiced skills and intuition. In addition to these direct effects, a number of secondary effects are likely. For example, passing on the wisdom of wayfinding from generation to generation is a meaningful opportunity for intergenerational exchange. In another context, Sheridan and Parasuraman [7] described the effects of "deskilling" and "being out-of-the-loop" when piloting highly automated aircrafts. On the one hand, they found that automation created security and safety, because the number of errors made by humans is reduced. On the other hand, the pilots are not able to further develop their skills of controlling a plane, in case the automation stops working.

Thus, while the desire to increase efficiency seems reasonable, we need to remind ourselves to question automation now and then because the "[…] ultimate purpose of technology is to make life better for people" ([7], p. 124)—not necessarily faster or simpler. Regarding the example of the toaster, it might be a valuable shortcut compared with roasting bread in an oven, but some may actually miss the replaced activity—for example, because they had been especially skillful bread roasters or for the smell of freshly roasted bread. They may have enjoyed the resulting morning ritual or loathe that the time supposedly saved now becomes occupied by the less inspiring task of cleaning the toaster. In this sense, whenever an everyday practice is further automated, this removes not only an unbeloved chore, but also a potential to involve oneself in an enjoyable activity. While the modernist narrative of the ever more powerful automation narrowly understands "better life" in terms of getting rid of supposedly unwanted chores, a more humanist perspective may question this oversimplification. To us, a "better life" does not necessarily imply more efficient technology, but focuses more on fostering people's well-being by designing enjoyable and meaningful everyday experiences [8]. Since the material (i.e., things, tools, devices, technology) inevitably shapes activities and experiences, the degree of automation we design into these things will inevitably impact the quality of resulting everyday experiences.

In the present paper, we argue that certain types of everyday automation unnecessarily streamline originally experientially rich activities, and thus, remove everyday enjoyable and meaningful moments. In two earlier studies, we explored how automation impacts the experience of coffee-making [9] and how automation and experience can be reconciled through a particular design strategy using the example of a grinder [10]. We will first provide a brief overview of the two previous laboratory studies (section 2). We then extend our finding to "the wild" (sections 3 and 4). Specifically, we provided people with a more manual alternative to automatic coffee brewing at home and explored to which degree the new practice was adopted and whether it had the expected positive impact (study 1).

In the same vein, we provided people with a prototypical semi-manual coffee grinder, which we call Hotzenplotz, to use at home as an alternative to their commonly used electrical grinder and explored resulting adoption, experience and meaning (study 2).

## 2 Previous work

### 2.1 Why coffee-making?

We chose the mundane activity of coffee-making (i.e., all steps from grinding through brewing) as our subject of study for a number of reasons (see also [11]): First, it is a common everyday activity at least in our immediate cultural environment. This eases the identification of participants, who practice the activity in everyday life. Second, it is an activity with a substantial heterogeneity, potentially involving a wide variety of different materials (e.g., coffeemakers, grinders) and meanings (e.g., being stimulated, rituals, togetherness). All in all, its cultural pervasiveness, the wide array of already available tools, machines, and techniques, and its apparent experiential qualities turn coffee-making into an ideal practice for our study interest. Interestingly, modern automated coffee machines embody exactly the opposite approach. The presumably meaningful elements, such as brewing the coffee and grinding the beans, are done by the machine, while the user is allocated cleaning and maintenance tasks (refilling water, emptying the drip tray).

### 2.2 The experiential costs of everyday automation

In a previous study [9], we took a closer look at differences between experiences emerging from a more "automated" coffee brewing (with a Senseo pad machine) and more "manual" coffee brewing (with a manual grinder in combination with an Italian coffee maker, Fig. 1). Obviously, what we labeled as "manual" also made use of tools, such as an Italian coffee maker or a grinder. However, compared with the fully automated Senseo pad machine, using the particular combination of tools afforded a variety of interactions, which required more manual input and provided more direct contact with coffee beans, powder, water, and heat.

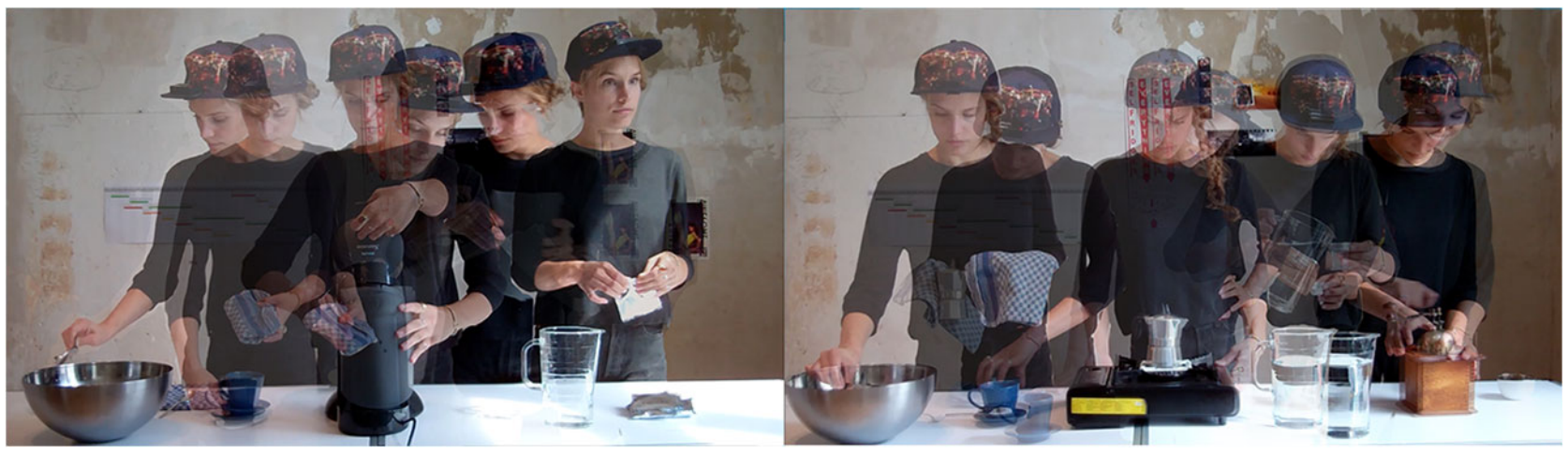

**Fig. 1** Automated (left) and manual setup (right)

Participants ($N = 20$) were invited to our laboratory and asked to prepare two cups of coffee in both the more automated and the more manual way (in a counterbalanced order). After each brewing process, they filled in a questionnaire measuring positive and negative affect as well as psychological need satisfaction (see section 3.2 for more details). Furthermore, they described their most positive and most negative moment during the activity. In addition, we measured the time it took to complete each brewing process (i.e., time on task).

Overall, brewing coffee manually was rated as more positive and more need fulfilling than brewing in an automated way. Feelings of competence and stimulation were significantly higher, when brewing more manually. However, the manual process also led to more negative affect, mostly due to problems during the process, e.g., because cranking was exhausting. In addition, manual brewing took significantly longer (manually 11:22 min versus automated 3:03 min on average). Positive moments in the more manual brewing were mainly related to the process itself (e.g., grinding), while in the more automated brewing the outcome, that is, drinking the coffee itself, had been the most positive moment. Interestingly, negative moments in the more manual condition were related to the process (i.e., mostly usability problems of the tools involved) and the outcome (i.e., disappointing result), while in more automated condition, 7 out of 20 participants complained about the waiting time. This was somewhat surprising, because the preparation time in the more automated condition was only about a third of the time used in the more manual condition.

Here, the dilemma of everyday automation becomes obvious: While automation is efficient and convenient, meaningful moments potentially derived from coffee-making are lost, since automation renders activities experientially "flat" in terms of need fulfillment and affect (positive as well as negative). With more manual brewing, participants enjoyed the smell and feel, that is, the sensory stimulation especially while grinding.

They felt as a part of the process through the transparency of each single step. They also felt in control and competent. While this engagement could lead to positive as well as negative moments, automation made the whole activity more or less disappear. In sum, this study showed that different ways of performing an activity (i.e., making coffee) leads to different experiential consequences. Automation is efficient but creates experiences of impatience, while a more manual process provides more positive affect and more need fulfillment, i.e., meaning. While from a convenience perspective, people who used the automated coffee maker saved about 8 min that could be invested in another, more enjoyable activity, from a well-being perspective the automated coffee maker took away an experientially satisfying everyday opportunity for well-being.

## 2.3 Reconciling automation with experience

The previous study demonstrated the differences in experiences emerging from different materials (i.e., tools, machines) as well as the different costs involved (e.g., time). Of course, people are free to choose the way to complete tasks in everyday life. There are times when people rely on automats, i.e., compressive designs, to enjoy an outcome (such as a cup of coffee), while actually having to do something else, which is more meaningful and enjoyable. In contrast, there are times when the activity itself is in the fore and more donative designs invite to spend time with the activity itself. This may create a dichotomy, where people engage in "slow" ways of doing certain activities on special occasions (e.g., on weekends, holidays), while they resort to less experientially satisfying ways of doing an activity in everyday life. However, many people already rely on everyday automation without having a more experientially satisfying way of doing an activity in their repertoire. In fact, in the study reported above [9], a number of participants had never prepared coffee with an Italian coffee maker. Thus, in everyday practice, these people cannot make an independent choice every time they make coffee, but the automats impose a certain experience or at least predefine coffee-making as an activity without any potential for enjoyment and meaning.

A possible solution is to reconcile automation with experience, that is, to identify ways to preserve certain advantages of automation, while providing a richer experience. In a further study [10], we deliberately designed an interaction for an automatic grinder to preserve experiential quality. The grinder *Hotzenplotz* [10] (Fig. 2) was based on a concept we dubbed

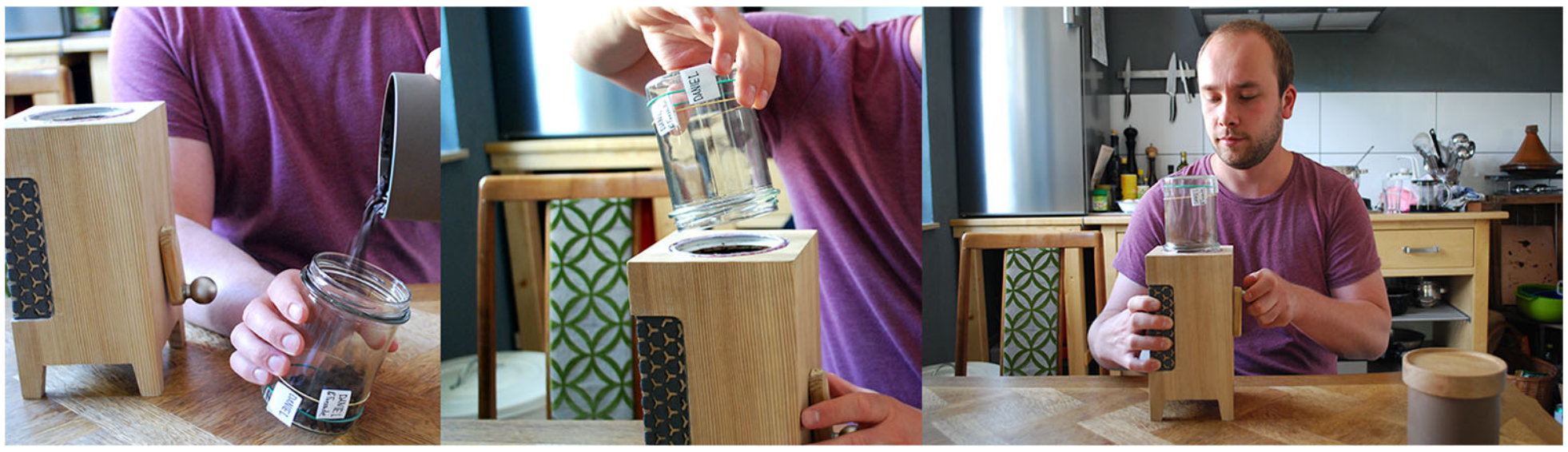

**Fig. 2** *Hotzenplotz* in action

"automation from below." Unlike conventional automation, where an entire activity is automated in a top-down manner, this kind of interaction starts from a manual interaction (to preserve meaning), which is then supported by automation. An everyday example would be an electric bike, which supports pedaling but does not replace it. In this study, we fitted an automatic grinder with a crank. A rotation sensor inside the crank detected turning, and according signals were forwarded to an Arduino board. A simple program controlled the speed of the motor of the grinder depending on the turning motion: Slow turning led to a slow rotation of the grinder's blades; fast turning accelerated the blades. In other words, *Hotzenplotz* borrowed a manual interaction to presumably inject meaning into the activity of switching on an automatic grinder. It replaced the push-button normally used to control electric grinders.

In this study, participants ($N = 15$) were invited to our laboratory and asked to grind coffee with three different grinders: a manual grinder, an electric grinder, and *Hotzenplotz* (in a counterbalanced order). After using each grinder, they filled in a number of questionnaires to measure positive and negative affect, need satisfaction, and hedonic and pragmatic quality (see section 3.2 for more details).

As expected, manual as well as combined *Hotzenplotz*-style grinding were experienced as more positive and more meaningful than electric (automated) grinding. Additionally, *Hotzenplotz* grinding outperformed manual grinding in its hedonic quality. Moreover, it was also perceived as more pragmatic than both the electric and the manual grinder. Concerning time perception, participants would have preferred electric grinding to be shorter and manual grinding was balanced ("about right"). With *Hotzenplotz*, however, participants wanted to prolong the interaction.

All in all, *Hotzenplotz*, our combined grinding device, had all the advantages of an electric grinder (e.g., finer powder, low effort) but created an experience significantly more positive and meaningful than the electric grinder. A simple change in interaction tremendously changed the experience of the same machine. *Hotzenplotz* was as good as the manual grinder on all measures, occasionally even slightly better. Contrary to the popular notion of "form follows function," a technically superfluous crank instilled meaning typically lost when using an automatic grinder. However, this technically unnecessary feature allowed for an interactive experience that combined and further improved the advantages of both the automated and manual alternatives in the lab. Similar strategies have also been helpful in making automated driving more meaningful and enjoyable [12].

Both previous studies showed that the way we perform everyday activities impacts tremendously the resulting experience. The material, that is, machines and tools, plays an important part in this. In the first study, we arranged a number of tools to shape the activity in either a more or less experiential way. In the second study, we successfully manipulated a technology in a way to instill more meaning. In both cases, the less automated alternative led to more positive experiences and, thus, more well-being [8]. However, whether this finding transfers to coffee-making in realistic everyday life scenarios remains to be answered. While participants in the laboratory may have time and the resources to engage in the activity and its reflection, this may neither be possible nor wanted in the real world. As a consequence, reshaping activities towards more well-being through the material may remain less successful, when used in everyday life. To explore this further, we conducted two field studies to complement the previous studies [9, 10].

## 3 Study 1: reshaping everyday experiences

To extend our findings from the previous lab studies, we ran a field study to test whether the experiential benefits promised by manual coffee brewing hold up in a real setting. We introduced 10 households, which typically prepare coffee automatically (i.e., with a fully automated coffee maker or an automated pour-over coffee maker that used pre-ground coffee beans, see Fig. 3a) to an additional, more manual way. Specifically, we were interested in whether households implement the alternative way of preparing coffee offered through the new materials given, as well as whether experiential advantages found in the laboratory also emerge under more realistic conditions and over time.

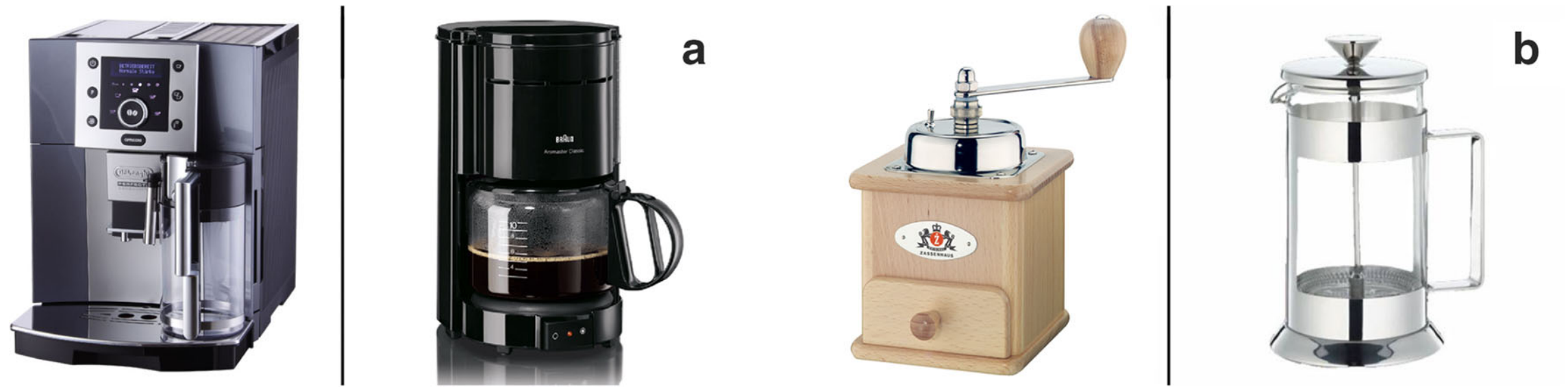

**Fig. 3** **a** De'Longhi Perfecta (left) and Braun Aromaster (right). **b** Zassenhaus Brasilia (left) and Cilio Laura (right)

## 3.1 Participants and procedure

Twenty-five individuals (see Table 1) from ten distinct households participated in the study (14 female, median age = 50, min 17, max = 77). They received no compensation for participation. The participants had diverse professional backgrounds, such as student, executive employer, or pensioner. All owned an automated coffee machine used as the primary way of preparing coffee in everyday life. The participants did not know the exact aim of the study. They were given the materials, asked to report about their coffee-making and try a new way of coffee-making. Participants accepted a data privacy declaration.

The study ran over 2 weeks. In the first week, our baseline condition, the participants were instructed to continue using their regular automated coffee maker (e.g., *De'Longhi Perfecta* or *Braun Aromaster*; see Fig. 3a) whenever they felt like having a coffee. We asked all households to fill in a short paper questionnaire in German after each time they made coffee to report about their experience. The questionnaire had a short mandatory section and an extended section to fill in on a voluntary basis. Each questionnaire was dropped into a box after it was filled in.

In the second week, the treatment condition, each household was provided with a kit, consisting of a bag of coffee beans, a French Press coffee maker (*Cilio Laura*), and a manual coffee grinder (*Zassenhaus Brasilia*; see Fig. 3b). Participants were instructed to use this kit at least once to prepare coffee instead of their regular machine. After that, it remained free to them to use either way of preparing coffee, whenever they felt like it.

Of course, participants with a fully automated coffee maker could use the provided coffee beans for the automated preparation as well. Participants were again asked to fill in the same questionnaire with a mandatory and an extended section after each coffee-making.

**Table 1** Background information of the participants (study 1)

| household | H1 | H2 | H3 | H4 | H5 | H6 | H7 | H8 | H9 | H10 |
|---|---|---|---|---|---|---|---|---|---|---|
| gender/ household size | | | | | | | | | | |
| age | 48, 49, 20, 17 | 54, 59, 23 | 58, 56 | 48, 74, 17, 24 | 48, 48, 17, 14 | 53, 58 | 57 | 77, 77 | 50, 50 | 60 |
| baseline-condition | fully-automated coffee maker | fully-automated coffee maker | fully-automated coffee maker | fully-automated coffee maker | fully-automated coffee maker | Pour-over Coffee maker | fully-automated coffee maker | Pour-over Coffee maker | Pour-over Coffee maker | fully-automated coffee maker |

## 3.2 Method: questionnaires

The mandatory section of the questionnaires consisted of a field to enter the exact date and time as well as an identifier for the person preparing coffee, the coffee machine they used (automated coffee maker or grinder and French press; this field was only included in the second week), the amount of cups they prepared, and for whom (for themselves, family members, and/or guests). This was followed by two items about how positive and negative participants felt during coffee-making (both on a 7-point Likert scale from 1 = "not at all" to 7 = "extremely") and four questions relating to their subjective focus with the same scales. With the focus questions, we asked them if they felt to be under time pressure, whether they could take the time they wanted, whether they felt as if this were having a break, and finally whether they could concentrate on the activity.

In the extended part, we included a version of the Positive Affect Negative Affect Schedule (PANAS, [13]) in a German translation [14] to measure positive and negative affect during coffee preparation in a more detailed way than the single items (see above). The PANAS consists of ten attributes in two scales measuring positive and negative affect during an activity, and participants were asked to answer whether they felt like that (here: during coffee preparation) on a five-point scale (1 = "not at all"; 2 = "a little"; 3 = "moderately"; 4 = "quite a bit"; 5 = "extremely"). The attributes on the positive affect scale are "interested," "excited," "strong," "inspired," "active," "jittery," "attentive," "determined," "enthusiastic," and "proud." For negative affect, the attributes are "disinterested," "upset," "guilty," "scared," "hostile," "irritable," "alert," "nervous," "afraid," and "ashamed."

Besides the PANAS, we quantified the experience by means of six "Need Fulfillment" scales [15] in a German translation based on Sheldon et al. [16]. These scales measure psychological need fulfillment with three items for each need. We included the six needs that had previously been identified as particularly relevant for activities including interactive technologies, namely autonomy, relatedness, competence, stimulation, popularity, and security, resulting in 18 items in total. Typical items were "While grinding, I felt that I was successfully completing a difficult task" (competence) or "While grinding, I felt glad that I have a comfortable set of routines and habits" (security). Responses were given on a five-point scale ranging from "not at all" (1) to "extremely" (5). Finally, we included four open questions asking for the most positive and most negative moment during coffee preparation, suggestions to improve the coffee preparation and other remarks. The open questions are the basis for the qualitative analysis reported in section 3.3. Internal consistencies of the focus scale, need scales, and the PANAS are reported in Table 2, all of which indicated a satisfactory reliability of our measures.

### 3.3 Quantitative findings

For our analysis, we aggregated the answers per participant and condition. For example, we calculated the mean value for all answers concerning need fulfillment during the baseline condition, and one each for the manual and automatic preparation in the second week. This is a rather conservative way to analyze differences between the three conditions (baseline, manual, automatic), as it reduces the degrees of freedom dramatically. However, it allowed us to calculate within-subjects tests.

#### 3.3.1 Frequency of use

In total, participants prepared coffee 472 times (286 times for themselves, 37 times for their family members, 10 times for guests, and 139 times where they prepared coffee for more than one of these categories, e.g., for themselves and for guests). They prepared coffee 273 times in the baseline condition (week 1) with 91 short and 182 extended answers, i.e., about four times per household and day. In the treatment condition (week 2), participants reported 199 instances of coffee-making (85 short and 114 extended answers), thus slightly less than in the baseline (about three times per household and day).

**Table 2** Internal consistencies ($N = 472$)

| Measure | Cronbach's α |
|---|---|
| Focus | .79 |
| Psychological needs (overall) | .97 |
| Autonomy | .81 |
| Competence | .89 |
| Relatedness | .93 |
| Stimulation | .87 |
| Security | .86 |
| Popularity | .92 |
| Positive affect | .90 |
| Negative affect | .73 |

The coffee-making in the treatment condition was almost equally distributed between the new manual setup (106 of 199, 53%; 60 extended answers) and the familiar automated setup (93 of 199, 47%; 53 extended answers). While the manual preparation did not entirely replace automatic preparation, it was nevertheless practiced in everyday life. In the treatment condition, six (of 25) participants prepared coffee manually only, three prepared coffee more often manually than automatic, seven prepared coffee more often with the automatic machine than manually, and six used the automatic machine only. Three participants prepared no coffee at all in the second week. Two of those were members of households in which other people commonly prepared coffee for family members (12 and 24 times), indicating that they might have received these. The other person was from a household where all members only rarely drank coffee at all (9 times in 2 weeks), but prepared most coffee for guests (11 times).

#### 3.3.2 Affect

Given the large number of extended replies, we report the results for the more established PANAS instead of the single question for affect. First, we calculated the difference between the positive and the negative affect scales (the averages of the respective ten items) as a compound measure for "affect balance." This compound measure runs from −4 to +4, with positive values indicating the predominance of positive affect and vice versa. We ran a one-way repeated measures ANOVA with the factor levels "baseline" (in week 1), "automatic coffee-making" (in week 2), and "manual coffee-making" (in week 2), and with affect balance as measure. We found a significant main effect, $F_{(2,\ 18)} = 10.55$, $p < .001$, $\eta^2_p = .54$. Bonferroni-corrected pairwise comparisons revealed a significant difference between the baseline ($M = 1.18$) and manual coffee-making ($M = 2.10$; $\delta = 0.91$; SE = 0.23; $p < .01$) and between automatic ($M = 1.11$) and manual coffee-making ($\delta = 0.98$; SE = 0.31; $p < .05$). The baseline condition and automatic coffee-making in week 2 did not differ ($\delta = 0.07$; SE = 0.15; $p = 1$). Taken together, we can conclude that manual coffee-making led to a more positive affective experience than both automatic coffee-making in the baseline condition and automatic coffee-making after the manual kit was introduced.

#### 3.3.3 Focus

We inverted the question about subjective time pressure and then aggregated the four focus questions by calculating the mean to measure "focus." To analyze differences in focus, we ran a one-way repeated measures ANOVA with the factor levels "baseline" (in week 1), "automatic coffee-making" (in week 2), and "manual coffee-making" (in week 2) with "focus" as our dependent measure and found a significant main effect, $F_{(2,\ 18)} = 4.54$, $p < .05$, $\eta^2_p = .34$. Again, we ran pairwise comparisons between the three factor levels. However, manual coffee-making ($M = 5.74$) did not differ significantly from the baseline ($M = 5.05$; $\delta = 0.69$; SE = 0.31; $p = .17$) after Bonferroni correction. Manual brewing also missed significance when compared with automatic coffee-making in week 2 ($M = 4.61$; $\delta = 1.13$; SE = 0.48; $p = .13$). No difference was found between the baseline and automatic brewing in week 2 ($\delta = 0.44$; SE = 0.31; $p = .56$). Taken together, we found a difference in focus between the three conditions, with the descriptively highest focus during manual coffee-making, followed by the baseline.

#### 3.3.4 Need satisfaction

Finally, we analyzed the results from the need satisfaction scales. First, we calculated need satisfaction as the mean of the three items per need, which served as our repeated measures. Then, we ran a 6 × 3 repeated measures ANOVA with the factor "needs" and the needs autonomy, relatedness, competence, stimulation, popularity, and security as the six factor levels. The other factor ("condition") consisted of our three conditions "baseline," "automatic coffee-making," (in week 2), and "manual coffee-making."

The ANOVA revealed a significant main effect for "needs" ($F_{(5,\ 30)} = 5.47, p < .001, \eta^2_p = .48$), and a main effect for "condition" ($F_{(2,\ 12)} = 7.01, p < .01, \eta^2_p = .54$). Moreover, the interaction effect was significant, $F_{(10,\ 60)} = 2.74, p < .01, \eta^2_p = .31$. We ran Bonferroni-corrected pairwise comparisons for the "condition" factor. Baseline ($M = 2.93$) and manual coffee-making ($M = 3.65$) did not differ significantly ($\delta = 0.72$; SE = 0.25; $p = .09$), nor did the baseline and automatic coffee-making in week 2 ($M = 2.94$; $\delta = 0.02$; SE = 0.19; $p =$ 1). The comparison between manual and automatic coffee-making just missed significance ($\delta = 0.71$; SE = 0.22; $p = .052$).

Based on the findings in [9], where manual coffee-making led to more intense experiences of competence and stimulation, we ran Bonferroni-corrected pairwise comparisons for these two factor levels. As far as competence is concerned, there was a significant difference between manual ($M = 3.54$) and automatic coffee-making in week 2 ($M = 2.68$; $\delta = 0.86$; SE = 0.24; $p < .05$). Manual coffee-making did not differ significantly from the baseline ($M = 2.70$; $\delta = 0.84$; SE = 0.35; $p = .16$) and the baseline did not differ from automatic coffee-making in week 2 ($\delta = 0.02$; SE = 0.25; $p = 1$). As far as stimulation is concerned, manual coffee-making ($M =$ 3.86), similarly, differed significantly from automatic coffee-making in week 2 ($M = 2.37$; $\delta = 1.50$; SE = 0.25; $p < .05$) but not from the baseline ($M = 2.39$; $\delta = 1.47$; SE = 0.49; $p = .07$). Baseline and automatic coffee-making in week 2 did not differ either ($\delta = 0.03$; SE = 0.21; $p = 1$). We also explored the other needs but found no difference between any condition for any need.

All in all, we found that the three conditions differed in terms of need satisfaction. More specifically, manual coffee-making addressed competence and stimulation more intensely than automatic coffee-making—but only, if there is an alternative as in week 2.

#### 3.3.5 Summary of quantitative results

In sum, we found manual coffee-making in a realistic setting to be associated with a more positive experience overall than automatic coffee-making. The conditions also differed in terms of focus on the activity, although we cannot conclusively attribute this to manual coffee brewing. Finally, need satisfaction was higher during manual coffee-making when compared with automatic coffee-making, particularly in terms of competence and stimulation.

### 3.4 Qualitative findings

To complement the quantitative results, we conducted a qualitative content analysis of the replies to the open questions available for each instance of coffee-making. The positive and negative moments were clustered inductively into categories. All in all, we collected 168 positive moments (automated = 119/manually = 49) and 133 negative moments (automated = 89/manually = 44).

We found that positive moments during automated coffee preparation were often related to the outcome, i.e., the cup of coffee (55 of 119, 46%). Anticipation (21 of 119, 18%) was related to the taste and quality of the coffee itself and the efficient process (10 of 119, 8%). Related to the process were 24% (28 of 119) of the positive moments. Among these, 17% (20 of 119) were about the smell of the coffee, 6% (7 of 119) about switching on the coffee machine, and 1% (1 of 119) about putting a cup in the machine. Only 4% (4 of 119) were related to the brewing itself, which could be distinguished in 3% (3 of 119) positive moments about enjoying to prepare coffee for somebody else and 1% (2 of 119) about experiencing a time out.

When preparing coffee manually, most positive moments related to the process (68%, 33 of 49). These were quite diverse: 25% (13 of 49) mentioned grinding the beans, 23% (12 of 49) the smell during the grinding and afterwards, 14% (7 of 49) pressing the stamp of the French Press, and 2% (1 of 49) mentioned brewing in general, boiling the water, and experimenting with the grinder. On the other hand, only 18% (9 of 49) mentioned the coffee itself. Given that the device was new to the participants, 10% (5 of 49) of the comments were related to the novelty of the manual brewing experience. Finally, 2% (1 of 49) were about preparing coffee for somebody else and 2% (1 of 49) of being reminded of something in the past. Taken together, these results show that the manual brewing focuses on the process, while the automated brewing is outcome-oriented. Moreover, the positive moments people experienced during the manual coffee brewing were more varied than with the automatic machine.

All of the negative moments of the automated preparation were related to the process. Furthermore, 64% (57 of 89) of the negative moments were experienced when the machine called the user for action: 36% (32 of 89) related to filling the water tank, 10% (9 of 89) were about refilling the coffee beans, 9% (8 of 89) about descaling the coffee machine, 7% (6 of 89) described the moment of emptying the coffee ground container, and 2% (2 of 89) were about cleaning the machine. Thus, users were often annoyed by service activities they had to do for the machine to work properly. In addition, 26 of 89 (29%) comments were related broadly to a lack of autonomy or control. Sixteen percent (14 of 89) were about the loudness during the grinding process, because the built-in grinder started at maximum speed and thereby the loudness turned on to a maximum as well. Another 13% (12 of 89) disliked the waiting time because they could not influence it. Apart from these two themes, 7% (6 of 89) of the comments were about general usability problems. Regarding the manual process, 48% (22 of 44) of the negative moments were about the additional effort participants had to put into it. 16% (7 of 44) related to cleaning, 11% (5 of 44) to turning the crank, 9% (4 of 44) to refilling coffee beans, 3% (2 of 44) to boiling water, and 9% (4 of 44) to additional effort in general. Another 27% (11 of 44) related to the duration of the coffee brewing as rather long. Moreover, 16% (7 of 44) were general usability problems (9%, 4 of 44).

### 3.5 Summary of study 1

In this study, the two ways of brewing coffee, automatic and manual, differed quite substantially. The negative aspects of one condition are the benefits of the other. Our added insights into the experiential benefits of manual coffee preparation can help to combine both processes in an efficient and meaningful way. The automated brewing process was more outcome-oriented, which is why users were searching for moments to enjoy during the brewing process, such as the smell of the coffee, or moments of anticipation by switching on the machine. Efficiency was important to these participants. In contrast, users disliked to be commissioned to fulfill service tasks for the machine, such as refilling water and beans or cleaning. Moreover, we could show that interaction control is important, which was especially clear in this case with the noisy grinder. When we introduced the manual alternative, the affective experience of automatic brewing became more negative than it was before and also compared with the manual process. The latter was generally more stimulating and provided an experience of competence.

The manual brewing process was very prominent, and grinding the coffee beans was one of the most satisfying sub-processes because the user was in close contact with the coffee beans and could experience the transformation into a cup of coffee with all senses. This is brought about by the smell of the ground coffee or by pressing the stamp of the French Press. All of these sub-processes were enjoyable and led to a more stimulating experience. But of course, those sub-processes take more time and effort until the coffee is finally prepared. It is thus slightly more difficult to be integrated into the daily routines of our participants. Therefore, while manual coffee-making is more joyful, it is also less efficient. Nevertheless, the fact that about half of all instances of coffee-making in the second week were manual coffee-making reveals a much greater potential for increasing well-being by introducing more experiential ways of performing an activity than typically assumed.

## 4 Study 2: reconciling automation with experience in an everyday context

So far, we outlined the trade-off between a satisfactory experience and efficiency that comes with deciding for or against automation. The *Hotzenplotz* grinder (section 2.3.) attempts to reconcile this apparent contradiction between meaning and efficiency. It is a hybrid to combine the best of both worlds. In study 2, we tested *Hotzenplotz* "in the wild." We sought to explore whether promising previous findings transfer to everyday coffee grinding in real life. Based on our previous results, we expected *Hotzenplotz* to provide a richer experience, specifically in terms of stimulation and competence. We also expected a more positive affective experience with *Hotzenplotz*. In the present study, we further complement quantitative findings with an extensive qualitative investigation. Through this mixed method approach, we provide not only a holistic view, but also a detailed perspective on experiences of using a hybrid coffee grinder in everyday life.

### 4.1 Participants and procedure

In this two-week study, the grinding process as a sub-process of "coffee-making" was subject of the investigation. Eight individuals participated in the study (6 female, median age = 30, min = 27, max = 50; see Table 3). They had diverse backgrounds, such as civil engineer, psychotherapist and being students. They received no compensation for participation. Moreover, the participants did not know the exact aims of the study. Before the study, participants signed a data privacy declaration. Each of them was provided with each of both grinders (*Hotzenplotz* and automatic grinder *TZS Millard*, see Fig. 4) for 1 week in a counterbalanced order. Besides the grinders, we provided the participants with coffee beans and a French Press coffee maker.

In each condition, participants were briefly introduced to the respective grinder (see Fig. 4) and the French Press coffee maker. We asked them to make coffee using the provided tools over the course of 1 week. After each interaction with the grinder, but before brewing the coffee with the coffee maker, the participants were instructed to fill in an online questionnaire via their smartphones.

## 4.2 Method

### 4.2.1 Questionnaires

The questionnaire contained two items about how positive and negative participants felt during coffee grinding (both on a 7-point Likert scale from 1 = "not at all" to 7 = "extremely"). Then, we measured focus on the task by asking in how far they were concentrated on the grinding process and included another question about whether they would have liked to prolong the process, both with the same 7-point scale. Finally, we measured psychological need satisfaction using the same scales as in study 1. Additionally, we measured subjective product quality using the AttrakDiff Mini [17]. This questionnaire assumes two different, broad categories of perceptions: *hedonic* and *pragmatic*. Pragmatic quality refers to a product's perceived potential to support particular "do-goals" (e.g., to make a telephone call). In contrast, hedonic quality refers to a product's potential to support pleasure in use and ownership, that is, the fulfillment of psychological needs (e.g., to be stimulated). The AttrakDiff Mini includes nine seven-point semantic differential items, four of which capture pragmatic quality: *simple–complicated*, *practical–impractical*, *predictable–unpredictable*, and *clearly structured–confusing*. Four more items captured hedonic quality: *stylish–tacky*, *premium–cheap*, *creative–unimaginative*, and *captivating–dull*. In addition, we measured general goodness (*good–bad*) and beauty (*ugly–beautiful*) with a single item each. Internal consistencies were all satisfactory (see Table 4).

**Table 3** Background information of the participants (study 2)

| participant | P1 | P2 | P3 | P4 | P5 | P6 | P7 | P8 |
|---|---|---|---|---|---|---|---|---|
| Gender | | | | | | | | |
| Age | 28 | 50 | 30 | 30 | 27 | 28 | 30 | 30 |
| academic background | no | no | yes | yes | yes | yes | yes | yes |
| size of household | 1 | 4 | 2 | 3 | 1 | 1 | 2 | 1 |
| expertise in coffee-making | expert | moderate | moderate | moderate | moderate | moderate | moderate | moderate |
| private coffee-making | manually (man.) | automated (auto.) | auto. | man. | auto. | auto. | auto. | auto. |

### 4.2.2 Interviews

After each week of using one of the two grinders, we carried out a semi-structured interview to get a deeper understanding how each grinder was experienced and integrated into daily routines. At the beginning of the interview, we asked how the participants experienced the grinder in general, and we particularly asked for positive and negative moments. Then we asked if they would have liked to prolong or shorten the interaction, or if it just felt right. To continue, we asked about the experiences of need satisfaction, such as competence, stimulation, and autonomy. Furthermore, we asked the participants to describe the interaction in detail. Finally, end, we asked about the meaning attributed to the grinding itself. All interviews were audio- and video-recorded and later transcribed.

## 4.3 Quantitative findings

- Similar to study 1, we used the aggregated means of all coffee grinding instances per participant and condition. For instance, if one participant used one of the grinders five times, we aggregated the five values from the 5 filled-in questionnaires into a single value that we used for our data analysis. In order to control for a novelty effect with *Hotzenplotz*, we excluded the first 3 days of the *Hotzenplotz* condition for all calculations except for the frequency of use. Two participants used *Hotzenplotz* only during the first 3 days and were thus excluded from these analyses.

### 4.3.1 Frequency of use

- In total, the eight participants grinded coffee beans 99 times over the 2 weeks, 40 times (40%) with *Hotzenplotz*. This amounts roughly to one grinding per day and per participant, slightly less with *Hotzenplotz* compared with that of the automatic grinder. We compared the frequency of use for both machines by running a two-way ANOVA (2 × 2) with the within-participants factor "grinder" and the between-subjects factor "order of usage." Our dependent variable was the number of grinding per condition. We found only a significant main effect for "grinder" ($F_{(1, 6)} = 10.46$, $p < .05$, $\eta 2_p = .64$), indicating that participants used the automatic grinder ($M = 7.37$; SE = 1.25) more often than *Hotzenplotz* ($M = 5.13$; SE = 1.19), independent of the "order of usage."

### 4.3.2 Affect

Unlike the previous study, we did not include the PANAS this time in order to keep the questionnaire reasonably short. Thus,

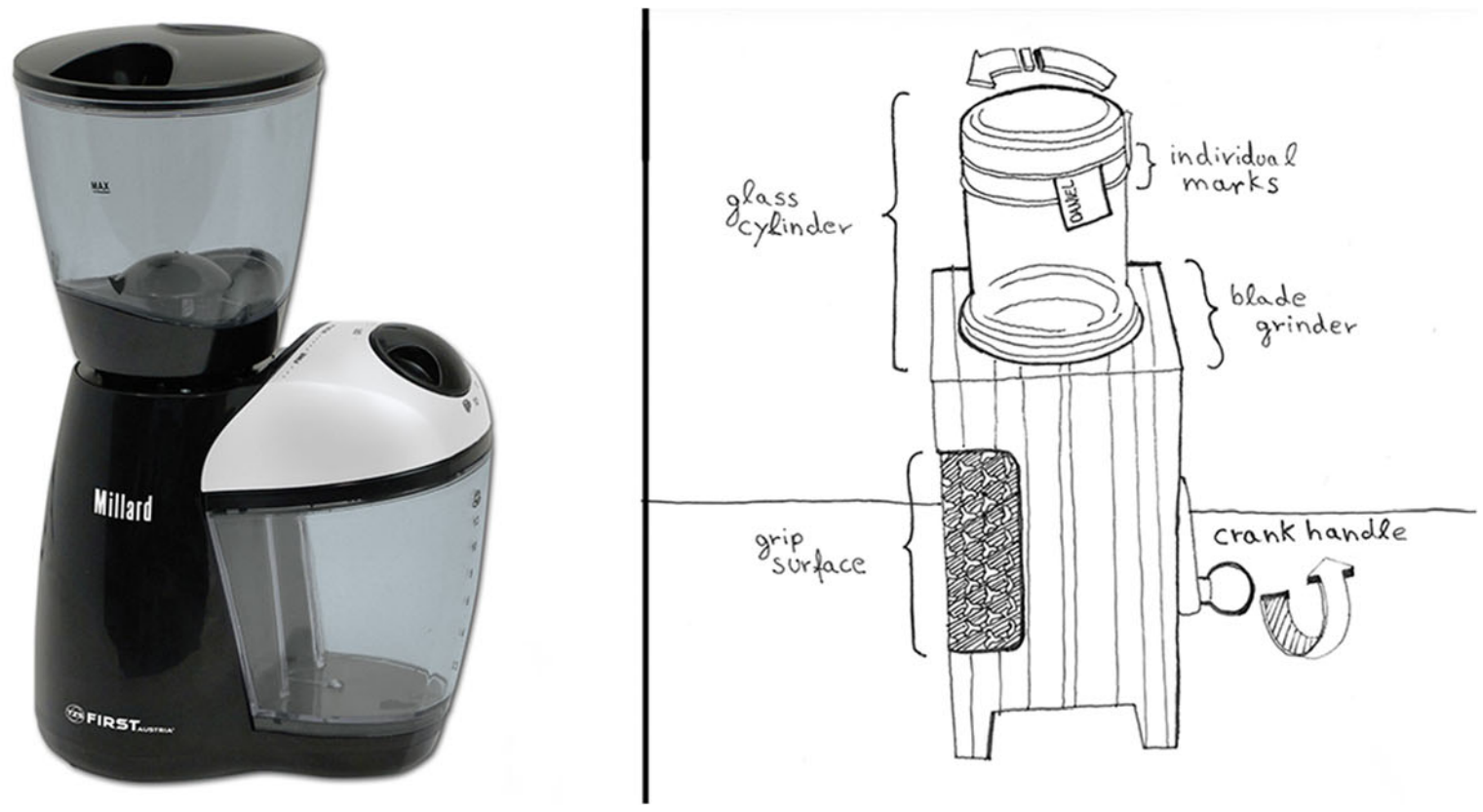

**Fig. 4** TZS First Austria Millard (left) and Hotzenplotz (right)

affect balance was calculated as the difference between the questions about positive and negative feelings during grinding. A 2 × 2 ANOVA with the within factor "grinder" and the between-factor "order of usage" revealed no significant main effect of "grinder" on affect balance, $F_{(1,\ 4)} = 1.84$, $p = .12$, $\eta^2_p = .32$. No other effect was significant.

#### 4.3.3 Focus

Next, we compared the effect of the grinding process on "focus." A similar 2 × 2 ANOVA with the measure "focus" revealed a significant main effect for the factor "grinder" ($F_{(1,\ 4)} = 18.81$, $p < .05$, $\eta^2_p = .83$), indicating that participants were more focused on the grinding process, when using the *Hotzenplotz* ($M = 4.97$; SE = 0.31), compared with the automatic machine ($M = 2.68$; SE = 0.57). No other effects were significant.

#### 4.3.4 Willingness to spend more time grinding

Moreover, we looked at the willingness to spend more time grinding coffee with the respective machines. Using the question whether participants would have rather lengthened the process as our measure, a similar 2 × 2 ANOVA revealed no effect for "grinder" ($F_{(1,\ 4)} = 3.38$, $p = .07$, $\eta^2_p = .46$), and no other effects were significant.

**Table 4** Internal consistencies of the scales ($N = 97$)

| Measure | Cronbach's α |
|---|---|
| Psychological needs (overall) | .95 |
| Autonomy | .88 |
| Competence | .77 |
| Relatedness | .93 |
| Stimulation | .91 |
| Security | .77 |
| Popularity | .88 |
| Hedonic quality | .93 |
| Pragmatic quality | .72 |

#### 4.3.5 Need fulfillment

Concerning need fulfillment, we ran a 6 × 2 × 2 mixed ANOVA with the within-factors "needs" and "grinder" and the between-factor "order of usage." We found a significant main effect for "grinder" ($F_{(1,\ 4)} = 7.76$; $p < .05$; $\eta^2_p = .66$), indicating that *Hotzenplotz* ($M = 3.24$; SE = 0.30) allowed for a more fulfilling experience than the automatic grinder ($M = 2.51$; SE = 0.25). Moreover, we found a main effect for "needs" ($F_{(5,\ 20)} = 4.66$; $p < .01$; $\eta^2_p = .54$) and an interaction effect of "grinder" and "needs" ($F_{(5,\ 20)} = 3.22$; $p < .05$; $\eta^2_p = .45$).

No other effect was significant. All needs were more fulfilled when using *Hotzenplotz* with the exception of security. Based on our previous findings in the lab, we ran two separate 2 × 2 ANOVAs with the factors "grinder" and "order of usage" for stimulation and competence, respectively. We found a significant main effect on the factor "grinder" for "stimulation," which was higher in the *Hotzenplotz* condition ($M = 3.77$; SE = 0.32) than in the automatic condition ($M = 2.48$; SE = 0.24; $F_{(1,\ 4)} = 10.77$; $p < .05$; $\eta^2_p = .73$). Moreover, we found a significant main effect for "competence," which was also higher in the *Hotzenplotz* condition ($M = 2.98$; SE = 0.38) than in the automatic condition ($M = 2.07$; SE = 0.17; $F_{(1,\ 4)} = 7.80$; $p < .05$; $\eta^2_p = .66$). Both effects confirm our previous findings from the lab. No other effects were significant in either ANOVA.

Similar to the previous lab study, interaction with *Hotzenplotz* was experiences as more fulfilling than the automatic grinder. Specifically, it was more stimulating and made the participants feel more competent.

#### 4.3.6 Product quality

The AttrakDiff Mini contains a scale for perceptions of hedonic quality, one scale for perceptions of pragmatic quality and one item each for general evaluation (i.e., goodness) and beauty. Concerning hedonic quality, a 2 × 2 ANOVA with the within factor "grinder" and the between-factor "order of usage" revealed a main effect for "grinder" ($F_{(1,\ 4)} = 17.18$, $p < .05$, $\eta^2_p = .81$), indicating that participants attributed a higher hedonic quality to *Hotzenplotz* ($M = 5.42$; SE = 0.48) than to the automatic grinder ($M = 3.08$; SE = 0.18). No other effect was significant. A similar 2 × 2 ANOVA for pragmatic quality revealed no difference between the two grinders ($F_{(1,\ 4)} = 0.83$, $p = .42$, $\eta^2_p = .17$), and no further effects. In terms of "goodness," a 2 × 2 ANOVA revealed a main effect for "grinder" ($F_{(1,\ 4)} = 8.47$; $p < .05$, $\eta^2_p = .68$). *Hotzenplotz* ($M = 5.94$; SE = 0.38) was perceived as better than the automatic grinder ($M = 5.09$; SE = 0.39). No other effect was significant. For "beauty," the assumption of normality was violated, which is why we opted for the non-parametric Wilcoxon test with the factor "grinder." *Hotzenplotz* was perceived as more beautiful (Mdn = 6) than the automatic grinder (Mdn = 3.17, $z = -1.99$; $p < .05$; $r = -.58$).

#### 4.3.7 Summary of quantitative findings

In sum, the quantitative data from study 2 revealed experiential advantages of *Hotzenplotz* when compared with the automatic grinder. These advantages stem from experiences of competence and stimulation. Although need fulfillment was higher, this was not mirrored in measured affect. We attribute this to the reduced affect scale and the small sample. However, hedonic quality, goodness, and beauty measures all indicated that *Hotzenplotz* was perceived more positive than the automatic grinder, with its pragmatic quality still intact. Thus, *Hotzenplotz* was successful in reconciling experience and pragmatic aspects, such as efficiency.

### 4.4 Qualitative findings

In general, within the *Hotzenplotz* setting, fewer negative experiences were mentioned (16 of 147, 11%) compared with positive ones (131 of 147, 89%). In the electric grinding condition, mentioned were more negative (78 of 144, 54%) than positive (66 of 144, 46%). For the remaining analysis, we focused on the positive experiences. Our analysis consisted of two elements to provide both, a holistic and an in-depth understanding of the interview data, that is, we looked at "both the forest and the trees" ([18], p. 223). We first focused on details, using the sub-coding method [19, 20], which was later followed by a synthesis and reflection.

#### 4.4.1 Positive experiences and meaning

We first structured our interview data using the sub-coding method [19, 20]. This starts with inductively creating top-level categories ("creation process," "grinder," "preferences") followed by a re-classification of the codes, if applicable, to more and more detailed sub-codes. This resulted in three "code-families," which we mapped in a hierarchical diagram, allowing for a structured overview (see Fig. 5).

At first glance, Fig. 5 shows that the *Hotzenplotz* grinder (yellow dots) is represented more frequently than automatic grinder (blue dots), due to the higher proportion of positive comments it received. In the following, we analyzed the three "parent categories" and the differential, underlying mechanisms for the two grinders following the broad structure of the hierarchy.

**"Creation process"** The "creation process," which represents the actual interaction with the grinder, i.e., to turn beans into coffee powder, was appreciated in both grinders (65 positive mentions with *Hotzenplotz*, 23 with the automatic grinder). In general, comments were more focused on the "interaction" than the "outcome" given the focus of our interviews. The participants enjoyed being part of the process with both grinders. While "transparency" (i.e., having an idea of what is going on in the grinder) was seen as about equally positive with both grinders, positive mentions about "control" (i.e., having an influence on the grinding process) were almost exclusively related to *Hotzenplotz*.

Looking at "transparency" in more detail, we found that sensory feedback, such as the smell, the sound of the beans hitting the glass cylinder, or just watching them turn into coffee powder were seen as positive. Also, haptic feedback, i.e., the vibration of the grinder, was perceived positively. P7 reported that she liked the experience with all her senses: "The sound is definitely nice. How the beans get smashed against the glass. I liked this […] it was fun to see the beans flying high and to see how they were turned into coffee powder."

In terms of "control," there were very few comments on the automatic grinder, essentially relating to portioning, i.e., being in charge of measuring and readjusting the amount of beans to be ground. In contrast, *Hotzenplotz* produced positive feedback in more diverse subcategories. Participants enjoyed being in control of the grinding speed, using the crank, and, similar to the automatic grinder, the portioning.

These facets of control were directly related to as well as enhanced by the transparency: Controlling the speed enabled the participants to control the loudness of grinding, a major negative point in study 1. Cranking gave them a direct connection to the grinding process, enabling them to feel the motion and to observe the beans transform into powder. Portioning, in both conditions, provided haptic and visual feedback and offered a certain autonomy. For the automatic grinder, one participant (P6) said: "I did [prepare] it in my own way, because I could decide on the amount [of coffee beans]." For *Hotzenplotz*, a representative comment was: (P2) "[…] it was important to me to control it [the grinding], probably that meaning is a bit too much, but [it was important for me] to know that every step is reasonable." All in all, by increasing control, our participants felt that they were part of the grinding process when using *Hotzenplotz*, even though they were supported by the semi-automated, effortless grinding interaction. This feeling was less prominent with the automatic grinder.

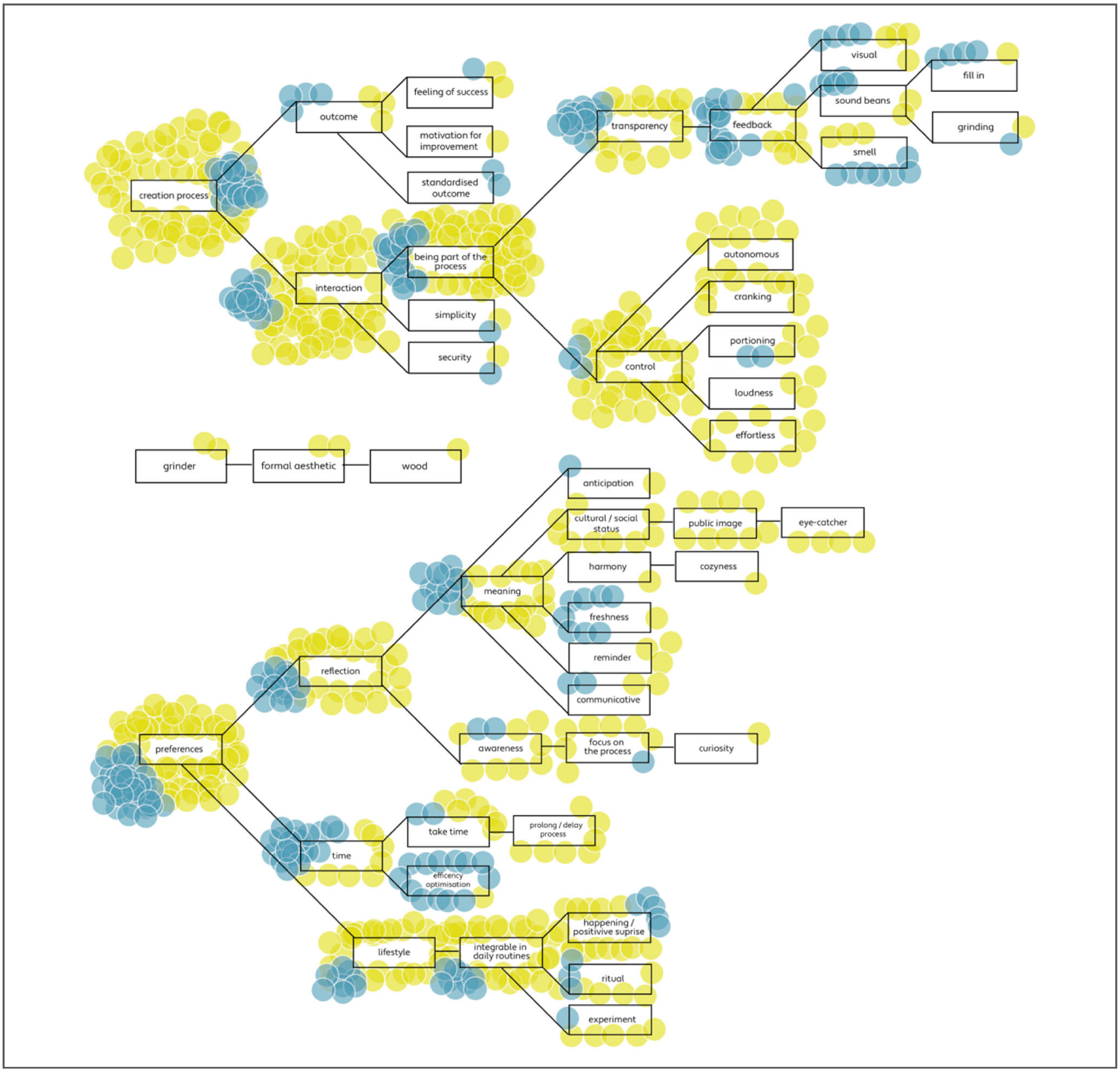

**Fig. 5** Code-landscaping: positive mentions related to the Hotzenplotz grinder (yellow dots) and automatic grinder (blue dots)

The outcome (fine coffee powder) was mentioned positively slightly more often with the automatic grinder and was mainly related to efficiency, since the participants could do something else in the meantime and were not focusing on the process as much.

**The "grinder"** The smallest coding family related to the form of the "grinder." Given that we asked mostly about the interaction, we had only a few positive comments about the device itself. A few participants reported the "formal esthetics" of *Hotzenplotz* as positive. P3 reported "regarding the optics, it was an eye-catcher for everyone entering my office"; P5 liked that *Hotzenplotz* suited the home context: "I liked about the machine that is was looking incredibly good, and it suits my kitchen in a nice way. I found this very positive." No participants reported positive statements about the formal esthetic of the automatic grinder.

**"User preferences"** The third coding family related to aspects of "user preference," i.e., how the grinder matched their overall lifestyle, how they reflected about using the grinder, and time-related attitudes. Generally, *Hotzenplotz* was positively mentioned in all three categories, while comments about the automatic grinder mainly related to time attitudes and reflection.

When users reflected about using the automatic grinder, they mainly thought about the freshness of the coffee powder. Unlike readily ground and packed coffee from the supermarket, the grinder allowed them to produce the coffee powder just before they made their coffee, which also triggered the anticipation of the coffee as a side effect. For instance, one participant said: (P4) "That was definitely part of the indulgence […] that's why I grind the coffee beans or let them be ground […] yes, it is also absolutely about the freshness. I mean freshly ground coffee is delicious." Moreover, *Hotzenplotz* facilitated communication with other people, such as guests, colleagues, or flat mates. As one participant put it: (P3) "Yes, one could easily start a conversation about the grinder." *Hotzenplotz* added further facets of meaning, such as cultural/social status through its character as an "eye-catcher," a raised awareness of ceremony for guests, or other people the coffee was prepared for. One person described it like this: (P3) "The rituals, which depend on daytimes, I have anyway, but this is an additional bonus, because coffee will be served and I made it." For some people it also triggered nostalgic feelings about their past, associated with harmony and coziness. Finally, the more varied interaction with *Hotzenplotz* served the participants' curiosity.

Although we had time-related comments for both grinders, the specifics were quite different. The automatic grinder elicited comments mainly about efficiency or optimization. For example, one participant said: (P5) "It is an operation you don't need to pay attention to. When you start it once you can do something else in parallel, because you don't need to take care of the machine. That's quite practical." Thus, it is valued particularly when time is scarce. In that regard, one participant (P7) stated "The electric one is very useful in the morning, because you will receive freshly ground coffee." The picture was very different for *Hotzenplotz*. Participants wanted to "take time" for the process and to have the positive experience last longer compared with the automatic grinder. One participant said: (P5) "One has to make sure to portion the right amount of coffee for a Bialetti, because it is just a small amount. I would love to prolong the interaction." And another one stated (P1): "Usually I thought I would love to crank longer, even if I did not need so much coffee, but that was really nice."

Finally, positive comments relating to "lifestyle" and the ability to integrate the grinder in one's personal daily routines were fewer for the automatic grinder than for *Hotzenplotz*. With the automatic grinder, positive experiences mainly related to the character of a "happening." As one participant put it: (P4) "Yes, it was still something special. At some point it might turn into an everyday thing, when I might think 'coffee-grinding – I don't mind' But after one week it was still a happening." A few comments also described it as some sort of a "ritual," especially in the morning, in spite of the stress. The automatic grinder could showcase its strengths as an efficiency-optimized device. One user said: (P4) "Of course it was fun, because everything worked out so fast. Actually, I didn't need to do anything […] I liked that I could do something else in the meantime." The efficiency might be a basic requirement to allow participants to squeeze their ritual in because it abbreviates it.

In contrast, the *Hotzenplotz* raised more diverse positive comments about the integration in daily routines. Mainly participants reported that the *Hotzenplotz* provides a magical moment during the grinding process, which leads to a "positive surprise." Some participants also reported that they took their time for a beloved "ritual" in the morning to start the day or in the afternoon to have a break. Through the various possibilities of changing the final coffee powder in its consistency, participants were motivated to reframe coffee grinding as an "experiment": (P2) "One might experiment a bit and what happened, might not be expected before, because this is not an "ordinary" coffee grinder and that is great fun. You look through into the glass cylinder and see how the coffee beans swirl around and you see how fine the powder already is during the grinding."

**Summary of qualitative findings** In the first part of the analysis, we presented the detailed characteristics of both grinders as experienced by the participants in three different coding families: the "creation process" is based on the interaction with the grinder during the grinding of the coffee beans, the coding family "formal esthetics" describes how the outer appearance leads to positive experience and as the last coding family the user's "preferences," which describe an individual way how the participants interacted with the grinder (see Table 5 for a further overview). All of those coding families had particular sub-practices, which we defined based on Shove et al.'s [21] social practice theory. Here, a social practice consists of three elements: The most abstract element is the meaning, which could be better described as a particular aim to perform the practice. Klapperich et al. [22, 23] define *meaning* further with the satisfaction of basic psychological needs such as stimulation, being competent, acting autonomous, feeling popular or having positive feelings of being secured by rituals. The performance of the practice is based on the practitioner's *competences*, which are needed to perform the practice. And those *competences* are always supported by a specific *material*, which could be a tool or an object. In Table 5, it is demonstrated how these three elements are interconnected with each other and if for instance the *material* is changed, the practice may also change through the transformation of the *competences* and the *meaning* of the practice.

The "creation process" consisted of positive experiences of the interaction and the outcome of the coffee grinding. In both conditions the sub-practice of "observing" and "portioning" occurred. Here, "observing" was in both ways connected with a similar *material*: the transparent bean-reservoir (automatic grinder) and the glass cylinder (*Hotzenplotz* grinder) allowed the user to understand what was going on inside the machine.

The automatic grinder gave an insight how the beans were transported to the grinding mechanism through the bean-reservoir. At the *Hotzenplotz* grinder, the glass cylinder allowed a view on the grinding mechanism itself and also showed in detail how the coffee beans were ground, which might be the reason for more transparency and a higher stimulating effect. The sub-practice of "portioning" was comparable in both conditions as well through a similar *material* (powder container and glass cylinder), which could be seen as an option to be in the loop of the process by using the *materials* as a tool to portion the right amount of coffee powder, which led to a feeling of being competent.

The main difference was the sub-practice of "cranking" during the grinding process with the *Hotzenplotz*. Here, an interplay between the desired fineness of coffee powder through the "observing" occurred while turning the crank; this made the participants part of the process and created a feeling of competence and autonomy.

To sum up, the automatic grinder was appreciated for its ability to create standardized high-quality coffee powder. Even though participants only played a limited role in grinding, they appreciated the smell and haptic experience (transparency). In addition, *Hotzenplotz* was experienced positively because it allowed for a greater degree of control. The interaction with the crank, control of the speed, and portioning were seen as positive.

**Table 5** Practices within the Hotzenplotz and automatic grinder condition

| automatic grinder | | | Hotzenplotz-grinder | | |
|---|---|---|---|---|---|
| **competencies/ activity** | **material** | **meaning/ psych. needs** | **competencies/ activity** | **material** | **meaning/ psych. needs** |
| **'creation-process'** | | | | | |
| **observing** | bean-reservoir | transparency, anticipation ***stimulation*** | **observing** | glass-cylinder | transparency, anticipation ***stimulation*** |
| **portioning** | powder container | being in the loop ***competence*** | **portioning** | glass-cylinder | being in the loop ***competence*** |
| | | | **cranking** | crank / glass-cylinder | part of the process ***autonomy, competence*** |
| **'formal-aesthetics'** | | | | | |
| | | | **presenting the grinder** | Valuable materials | Inspire others ***popularity*** |
| **'preferences'** | | | | | |
| **Efficient brewing /** save time | high speed grinder | having a ritual ***efficiency, security*** | **Ceremony/** take time | glass-cylinder / crank | having a ritual ***security*** |
| **valuing raw-material** | bean-reservoir | having a choice ***autonomy*** | **valuing raw-material** | glass-cylinder | having a choice ***autonomy*** |
| | | | **being curious** | glass-cylinder | being in the loop ***stimulation*** |
| | | | **positive surprise** | glass-cylinder / crank | discovering ***stimulation*** |
| | | | **experimenting** | crank / glass-cylinder | grow knowledge ***autonomy*** |

Regarding the formal esthetics, the unusual form of *Hotzenplotz* left a positive impression with its users and integrated well in the environment. There were no particularly positive mentions about the form of the automatic grinder. Here the sub-practice of "presenting the grinder" occurred, which was connected to the *material*, by using valuable and natural *materials* and created a sense of popularity by inspiring others.

Differences in lifestyle or usage scenarios were captured in the "user preferences" coding family. Both devices had their respective advantages depending on the context. Participants reflected on the automatic grinder as deriving its *meaning* by making fresh coffee beans easily accessible without needing much care otherwise. We described the sub-practice as "efficient brewing," which could save time. Here, the *material* is a high-speed grinder which saves time and could be thereby filled in compressive designs. This kind of efficiency could be seen as a basic condition to have a beloved coffee ritual in the morning, which creates a meaningful moment through security. In a similar vein, the *Hotzenplotz* created a sub-practice, which we called "ceremony": The *material* of the practice (glass cylinder/crank) is designed to make time to celebrate this delightful moment; the *meaning* behind this is to take time for having a coffee ritual which is making users feel secure. "Valuing raw-material" as a sub-practice occurred in both conditions through the similar process, of choosing the coffee beans autonomously and filling them into the grinder and then not losing the connection with the *material* until further processing.

In contrast, the *Hotzenplotz* grinder created more diverse sub-practices through its specific design: the grinder motivates the user to "be curious" and start "experimenting" with the *Hotzenplotz*, because of the combination of the glass cylinder, and the crank the user always has the feeling of being in the loop, which helps to support an autonomous handling, but also to grow knowledge about coffee grinding itself. Participants also often reported the practice "positive surprise," because they still had something to discover, which made them feel more stimulated.

In conclusion, both grinders served as a topic of conversation. *Hotzenplotz* represented a certain nostalgia, while being a unique eye-catcher that was associated with social status. Some users were willing to spend more time on the *Hotzenplotz* grinding process, because it was meaningful to them and it became a joyful moment in their daily routine.

### 4.4.2 Synthesis and reflection of study 2

For this section, we reintegrate the detailed facets within the coding families ("creation process," "grinder," and "user preferences") to draw a picture of how these details blend together in overarching scenarios. As a compound concept, social practices can be designed through intentional interconnections of their three elements: *material, meaning*, and *competences* [21, 23]. We summarize and outline somewhat idealized practices for both modes of coffee-making, integrating the findings from the previous analysis, before briefly reflecting on each.

**Integration through efficiency** Coffee is an important ingredient in the daily routine of busy workers: It literally fuels the knowledge economy. While people appreciate the good taste and high quality of freshly ground beans, they have little time in the morning before leaving the house, often taking their coffee with them. Efficiency is key and the coffee grinder needs to fit in optimized, condensed morning routines. The coffee itself is important, which is reflected in the number of comments on smell and the standardized, reliably good coffee. Additional cues such as the sound of grinding beans and visual feedback, as well as the possibility to portion just the right amount of beans, serve as an ambient assurance that everything is going as planned and underlines the positive anticipation of a tasty coffee, while the thoughts of the busy user are already focused on other, more important things. The grinder itself stands back; it is less important than the coffee it produces. Its visual appearance is therefore less important.

Later, on the way to work, people enjoy the coffee, appreciate its freshness, and take some pride in still having at least contributed al little to its production.

While this scenario seems to be already covered by fully automated coffee makers, we found that a light, manual gesture, such as the portioning in our case, can provide additional value. We learned from the interviews that users appreciated this basic element of control, even if they were focused on efficiency. We think that a concept of "ambient" coffee-making that happens in the background, but is not completely hidden, opens up an interesting field of development, that may focus on sensory feedback supporting appreciation of the coffee while leaving the user with the freedom to take care of other things.

**Integration through need satisfaction** Coffee also represents leisure and social pastime. When taking a break, or as a ritualized practice in the afternoon, the competent home barista needs full control of each step in the preparation process, but does not want to be bothered with unnecessarily exhausting interactions that only distract from the important things. *Hotzenplotz* provides direct control of the mill with a naturalistic crank, which allows for perfect, "handmade" coffee powder and augments the interaction in a nearly magical way. While grinding, visual cues of the fineness of the powder, its smell, and the haptic feedback of the supporting engine are necessary to exercise that control and offer the stimulating experience that turns a user into a barista. All of this takes place before the coffee is ready for the next step. At times, one of the biggest challenges during grinding for the users is to let go of the enjoyable activity when the coffee is right, because they are so focused on the process and tend to forget the importance of the outcome. Of course, when friends come to visit, they expect nothing less than the perfect blend from their personal coffee expert, and are charmed by the mysterious, eye-catching grinder that expresses value and high quality. When finally drinking the coffee, the curious device often serves as a starting point for engaging conversations.

On the other side of the scale, traditional manual grinders occupy the space. In our field study, however, we found that the "automated from the bottom" *Hotzenplotz* augments this experience by reducing effort, while leaving or even extending the sensory experience. We learned that cautious automation of unpleasant sub-steps, such as turning the grinder, while preserving or even extending the positive ones, e.g., through visual feedback, created positive experiences that transferred to the participants' everyday lives. The direct contrast with an automated grinder invited them to think more consciously about their coffee routines.

## 5 General discussion

Two studies showed how different degrees of automation in coffee-making can lead to a higher psychological need fulfillment and positive affect in realistic, everyday contexts. In both studies, the more manual practices with lower degrees of automation led to experiences of higher stimulation for the users and made them feel competent. They were more focused on the process and found the manual interaction more positive (study 1) than the automated coffee-making. Our qualitative analyses further informed these findings, by explaining how exactly the materiality of the devices (visibility of the beans, crank), their sensory perception (noise, vibration), and mode of interaction influence the users' experiences and determine the niche in which they can establish a positive practice in everyday routines.

In the case of coffee-making, two separate modes emerged and different degrees of automation cater for them. The efficiency-oriented practices ("compressive design") profit from more autonomous devices that work in the background. In contrast, slow practices during coffee breaks ("donative design") were experientially richer and better catered for by our "automated from the bottom" grinder. These two approaches can be transferred to other contexts, and have, for example, already been preliminarily adapted to autonomous driving [12].

Further explorations of different gradients of automation may better reconcile both concepts. For instance, Shove et al. [3] describe how the time and attention daily activities need can be assigned more flexibly through automation. In a similiar vein Silverstone [24] argues to raise a sensitive awareness in the matter how new technologies (e.g. media) could change daily routines. In the context of coffee-making, we found this when contrasting the automated grinder and *Hotzenplotz* in study 2. However, these modes are not necessarily distinct. A possible reconciliation would be a grinder that is "trained" by the user or memorizes the exact pattern of grinding, and thus provide a controlled but efficient and time-shifted grinding practice.

Of course, both studies come with limitations. Generally, not the complete experience with both grinders could be examined, and our analysis was limited to the (still relatively non-invasive) questionnaires and relied on retrospective interviews. Furthermore, we only had one prototype of the *Hotzenplotz* grinder, which made it hard to have a larger investigation with more participants. In field studies, there are also more external, uncontrollable influences, which take effect on the study results. However, the consistency in our findings between the previous lab studies (section 2) and the herein presented field studies (sections 3 and 4) appear promising evidence for a certain stability of the reported effects.

We believe that, as a metaphor, the holistic experience of a symphony orchestra in concert, despite its archaic and unfamiliar but somehow magical character, may be useful: The conductor controls the different parts of the orchestra by waving a baton through the air (in social practice terms: material). Having a distinguished sense for the music and grown knowledge over time (competence), the whole ensemble is only complemented by a passion for music, providing meaning. While not accessible to everyone, we can imagine that this supposedly old-fashioned practice provides an experience on a different level than merely listening to the same piece, even if played by the most distinguished artists, on a record. The user needs to experience the automated aspects of the process as an enhancement. Kaerlein critically describes the role of automation in form of a conductor or an illusionist. On the one hand, he argues "[…] the actions as a conductor are limited to a form of initiating and arranging" ([25], p. 156). And on the other hand, he pointed out that, depending on the level of automation, the user acts as an illusionist and the interaction reveals as a kind of theater. So, we as designers or developers of automated systems need to recognize that there is always the risk to manipulate the user in a negative way. From our perspective, design of automation—especially for the private domain—must always remain connected to the aim of subjective well-being on behalf of any potential person coming in contact with the automat.

Finally, what does our study tell us about automation? We think that we could make a point that takes "full automation" from being the modernist ultimate goal of the design of everyday utilities in our series of four studies, each of which showed that more manual practices have their experiential advantages. But rather than condemning either way, we believe that a strategy as described by Friedman and Yoo [26] may be helpful. They argue to pause and restart (design-) processes and reflect on how this could lead to meaningful interactions. Following this route, we hope to restart the discussion on levels of automation, this time with a focus on the subjective experience of the users, rather than technological finesse and unquestioned efficiency.

**Acknowledgments** We like to thank all the participants of the four studies, who were providing us with insights for the presented research.

**Funding** Open Access funding enabled and organized by Projekt DEAL.

## Compliance with ethical standards

**Conflict of interest** The authors declare that they have no conflict of interest.